# ARTICLE

# Atomic-scale observation of ordered structure induced by surface segregation in annealed Pt@Co core-shell nanoparticles

Kohei Aso,*[a] Hirokazu Kobayashi,[b] Shotaro Yoshimaru,[c] Miho Yamauchi,[c,d] Syo Matsumura,[e,f] and Yoshifumi Oshima[a]

The ordered structure of binary alloy nanoparticles determines their magnetic and catalytic characteristics. In the alloys after annealing, one of the components preferentially segregates on the surface to reduce surface energy. This surface segregation has been known as a factor in the construction of an ordered phase near the surface. However, the segregation-induced ordering has not been observed for nanoparticles. Here, platinum@cobalt (Pt@Co) core-shell nanoparticles were synthesized, and their structural changes after annealing at 600°C, 700°C, and 800°C for 3 hours were observed by a scanning transmission electron microscope. We discovered an $L1_0$–PtCo structure near the surface at 700°C, which was unexpected given the initial Pt:Co ratio of about 4:1. The $L1_0$–PtCo structure was considered to form due to surface segregation of Pt atoms and diffusion insufficient to mix Pt and Co atoms in the particle overall because the structure did not form at 600°C and 800°C.

## Introduction

Binary alloy nanoparticles have been attractive to improve the performance of fuel cell catalysis[1–4] and magnetic devices.[5,6] To achieve higher performance, researchers are attracted to control the ordering of the alloy structure.[7] In most cases, binary nanoparticles have been chemically synthesized, and such the as-grown nanoparticles are annealed to get ordered structure. Platinum–cobalt (Pt–Co) alloy nanoparticles are one of the promising materials because the particles show oxygen reduction reaction with a higher rate or ferromagnetism with a high magnetic moment, which is related to the ordered structure. In the phase diagram, two ordered structures are known to be formed.[8,9] $L1_2$-$Pt_3Co$ structure, in which Pt atoms occupy face-centered sites in face-centered cubic (fcc) lattice, is formed below the temperature of about 760°C when the compositional ratio of Pt atoms is about 70%–80%. When the ratio of Pt atoms is between 40%–60%, the $L1_0$-PtCo is generated, in which Pt(001) and Co(001) layers are stacked alternately along the *c*-axis at temperatures below 820°C. Of these two structures, $L1_0$ structure shows the higher catalytic reaction and ferromagnetic property.

The structure of nanoalloys has been shown not to consistently follow such a phase diagram due to surface effects. When annealing solid solution materials composed of different elements, one of the constituent elements may preferentially segregate on a surface, a phenomenon known as surface segregation. The surface segregation occurs because the surface energy of facets composed of one element is lower than that of the other element. Surface segregation affects the atomic arrangement of the layers below the top surface.[10–15] Pt atoms have been observed to preferentially segregate at the surface of the Pt–Co thin film to form the outermost surface layer.[10] For the (100)[11–13] or (111)[14,15] surfaces, researchers have reported that the Pt and Co layers are alternately formed in about 2 to 10 layers from the surface so that the Co layer is formed under the Pt surface layer. This Pt and Co layer stacking is comparable to the $L1_0$-PtCo structure. The result shows that Pt–Co nanoparticles may have the $L1_0$ structure of Pt–Co under specific annealing conditions, regardless of the particle's overall composition. These facts suggest that surface segregation can regulate the ordered structure of alloy nanoparticles. However, such an ordering has not been observed in the previous researches about Pt–Co solid solution particles.[3,16]

Here, by using a scanning transmission electron microscope (STEM), we investigated the structural evolution of Pt@Co core-shell nanoparticles (with a size of ~15 nm) at different annealing temperatures. We found that the $L1_0$-PtCo phase was formed by annealing at 700°C for three hours even in nanoparticles with 80% Pt atoms.

[a.] *School of Materials Science, Japan Advanced Institute of Science and Technology, 1-1 Asahidai, Nomi, Ishikawa 923-1292, Japan. E-mail: aso@jaist.ac.jp*
[b.] *Division of Chemistry, Graduate School of Science, Kyoto University, Kitashirakawa Oiwake-cho, Sakyo-ku, Kyoto 606-8502, Japan*
[c.] *Department of Chemistry, Faculty of Science, Kyushu University, 744 Moto-oka, Nishi-ku, Fukuoka 819-0395, Japan*
[d.] *International Institute for Carbon-Neutral Energy Research (WPI-I2CNER), Kyushu University, 744 Moto-oka, Nishi-ku, Fukuoka 819-0395, Japan*
[e.] *Department of Applied Quantum Physics and Nuclear Engineering, Graduate School of Engineering, Kyushu University, Moto-oka 744, Nishi-ku, Fuku*
[f.] *The Ultramicroscopy Research Center, Kyushu University, Moto-oka 744, Nishi-ku, Fukuoka, 819-0395, Japan*

† Electronic Supplementary Information (ESI) available: Additional description about the details of the experimental and calculation conditions, a table of diffusion parameters, five figures of S/TEM observation results, and a figure of diffusion length calculation. See DOI: 10.1039/x0xx00000x

## Experimental

Pt@Co core-shell nanoparticles were prepared by chemical synthesis.[17] Pt@Co core-shell structure is expected to be formed because the Pt atoms can form a nucleus faster owing to the faster reduction speed than the Co atoms. When the nuclei combine to form a single crystal particle, the <111> direction grows quicker than the other directions.[18] The as-grown particle is expected to have a concave cube shape that corresponds to a cube with six (100) facets and elongated corners toward the diagonal direction of <111>. A hexane solution in which the as-grown particles were dispersed was dropped onto a thin carbon film supported by a molybdenum (Mo) grid. To prepare annealed samples, as-grown particles were dropped on three different Mo grids, and the grids were annealed at 600°C, 700°C, and 800°C, respectively. The sample temperature was raised to each target temperature at a rate of ~ 0.6°C/sec before being annealed for 3 hours in a vacuum (~ 1 × $10^{-4}$ Pa). The samples were examined after they had been cooled to room temperature.

The obtained nanoparticles were analyzed by STEM in high-angle annular dark-field (HAADF) mode for atom-resolved imaging (ESI Fig. S1†) and energy-dispersive X-ray spectroscopy (EDS) mode for chemical mapping (ESI Fig. S2†). To increase the signal-to-noise (S/N) ratio, the images and maps are combined for various frames of the same area.[19,20] Because the nanoparticles were only about 10 nm thick, the signal counts of characteristic X-rays did not appear to be sufficient to obtain elemental mappings at the sub-nanometer scale. Meanwhile, when raising the probe current or the irradiation time to enhance the S/N ratio, the nanoparticles would be damaged by electron beam irradiation. In this study, the S/N ratio was improved by applying the data-driven denoising method, block-matching and 3D (BM3D),[21,22] to the original Pt and Co elemental mappings to identify the surface Co layer with a few nanometer thicknesses. More details of experimental conditions, such as sample preparations and acquisition conditions, are described in ESI†.

## Results and discussion

The as-grown Pt@Co core-shell nanoparticle was confirmed to have the concave cube shape since the four corners of Figure 1a were thick, while the center was thin. The EDS maps confirmed that the core and shell are made up of Pt and Co atoms, respectively (Figs. 1b–d). The compositional profiles of Pt and Co, obtained by averaging these EDS intensities along the vertical direction in the rectangle region of Fig. 1d, show that a few layers at the surface are composed of close to 100% Co atoms. The Co ratio gradually decreased as they entered the inside of the particle, and the Pt and Co ratios are about 90% and 10%, respectively, inside by 2.5 nm or farther from the surface (Fig. 1e). To confirm the compositional ratio of Pt and Co atoms in the nanoparticle, we made a three-dimensional volumetric model (Fig. 1g). The model was constructed to have the concave cube shape, compositional ratio of Pt:Co = 4:1 for the overall, Pt@Co core-shell configuration, and surface Co layers of 0.5 nm in thickness. We obtained the composition profiles of the model as shown in Fig. 1f, with a similar procedure to the experimental ones. Because the computed profiles replicated the experimental ones, the experimental particle also seemed suitable to think the compositional ratio of Pt:Co = 4:1 with Pt@Co core-shell configuration.

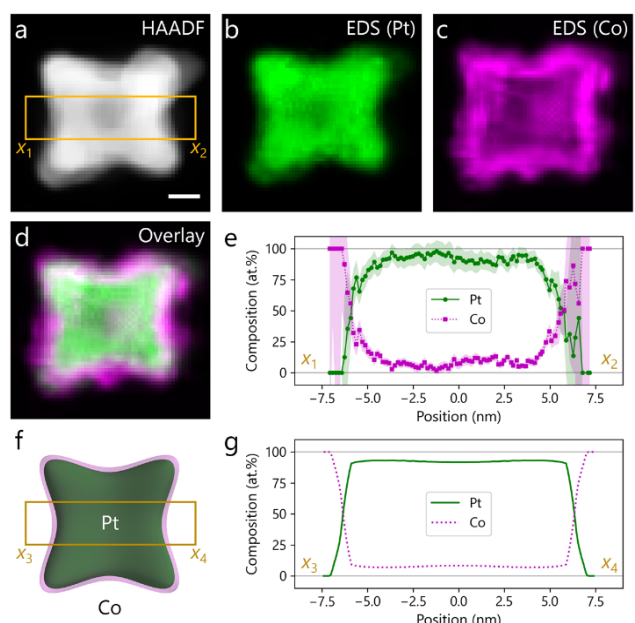

Fig. 1 Analysis of a typical as-grown nanoparticle. (a) HAADF image. The scale bar corresponds to 3 nm. EDS maps of (b) Pt, (c) Co, and (d) overlaid. (e) The composition profiles of Pt and Co as a function of the horizontal position on the line connecting $x_1$ and $x_2$. The EDS intensities of Pt and Co were averaged along the vertical direction in the rectangle region as shown in (d) and the averaged intensity profiles were converted into the composition profiles of Pt and Co. (f) Three-dimensional volumetric model of a Pt@Co core-shell nanoparticle. (g) The composition profile of the volumetric model.

The structures in the Pt@Co nanoparticles changed depending on the annealing temperature as shown in Fig. 2. The as-grown nanoparticle is again confirmed to have the concave cube shape with six {001} planes elongated in the outward <111> direction at each corner. The shape can be explained by the above-mentioned tendency to expand faster along the <111> direction than along the other directions.[18] The $L1_2$-$Pt_3Co$ structure was observed to form as illustrated by the green square in Fig. 2a, because the column intensity was lowest at the center site of the square lattice. Co and Pt atoms seemed to be mixed randomly around the interface of the core and shell. Given that the variation in column intensity between sites is visible weekly and locally, the $L1_2$-$Pt_3Co$ structure is most likely generated only in the limited region along the depth direction (red square in Fig. 2a).

When annealing at 600°C for 3 hours, the nanoparticles changed to the rounded cube shape as shown in the HAADF image of Fig. 2b. The shape change suggests the progress of the diffusion of atoms. In the image, the periodic square pattern, in which the darker atomic column is surrounded by brighter atomic columns, is observed to be arranged locally from the surface to about 2 nm. This contrast indicates the growth of the $L1_2$-$Pt_3Co$ structure. Because the square pattern is not always precisely aligned, various anti-phase boundaries are seen. The image contrast suggests that the $L1_2$-$Pt_3Co$ domains are grown from many different growth nuclei around the core-shell interface in the as-grown state. This structural change can be realized by mixing Pt and Co atoms. We can imagine that the diffusion of Pt and Co occurs during annealing, but the diffusion is not enough to create the ordered domains for a large area.

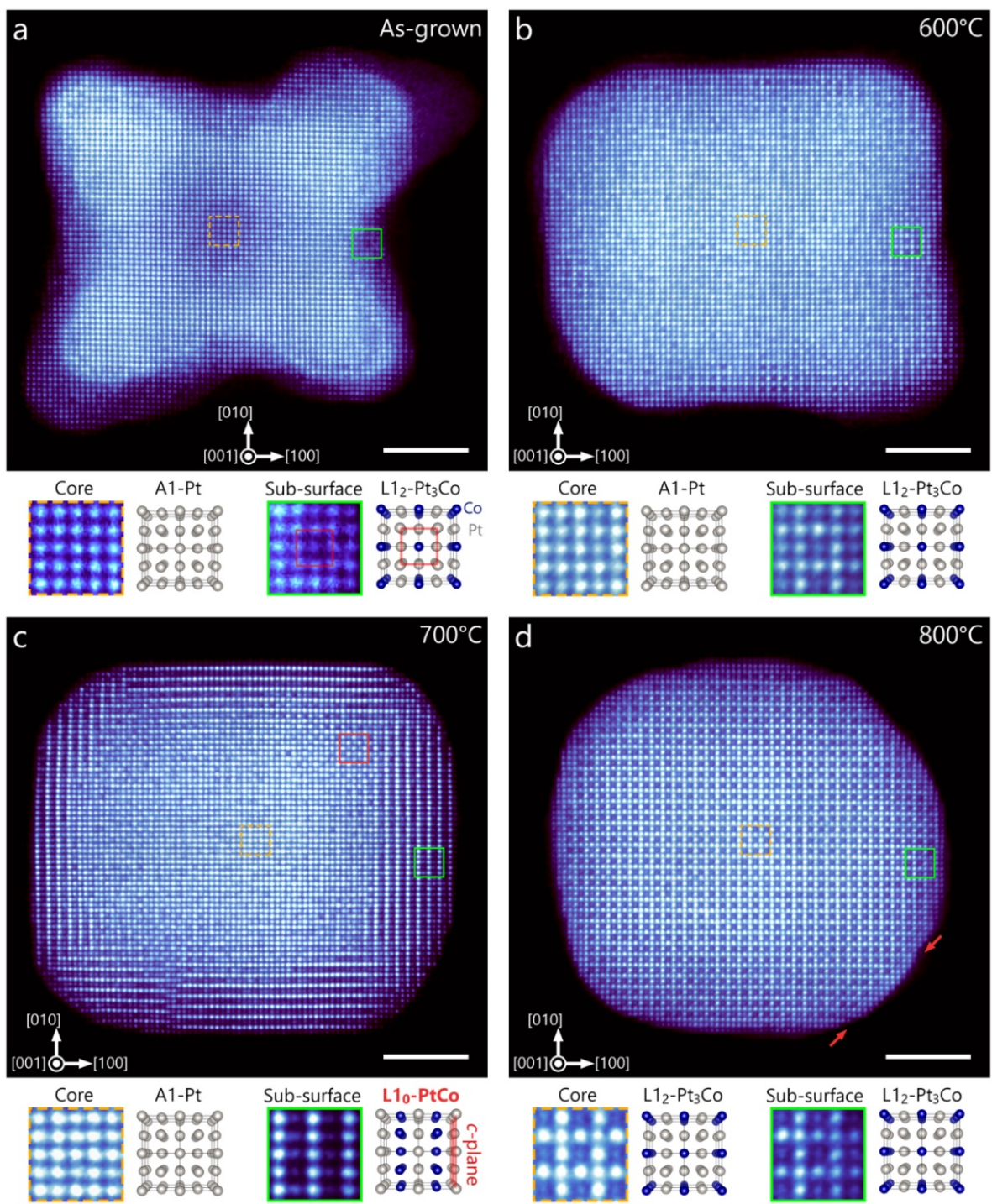

**Fig. 2** Atom-resolved HAADF images of nanoparticles (a) as-grown, annealed at (b) 600°C, (c) 700°C, and (d) 800°C. The scale bars correspond to 3 nm. Each lower left panel shows the enlarged views of the core (at the rectangle with dashed orange line) and sub-surface (at the rectangle with green line) with their corresponding atomic models. The red rectangle in c marks the checkerboard-like contrast of the $L1_0$ structure whose *c*-axis is parallel to the viewing direction. The red arrows in d mark the notable area of Pt surface segregation.

After 3 hours of annealing at 700°C, the nanoparticles were observed to be surrounded by flat {100} facets as displayed in the HAADF image of Fig. 2c. The morphological change occurs to create the stable low-index surface as the diffusion is more proceeded. At the region from the surface to about 2 nm, the luminous and dark layers are alternately stacked along the perpendicular direction to the surface. This alternate stack shows the $L1_0$-PtCo structure viewed along the *c*-plane. Although anti-phase boundaries are observed in some places, each $L1_0$-PtCo structure forms a large single domain. In the upper right corner of the nanoparticle in Fig. 2c, we observe that bright and dark spots are arranged alternately like a checkerboard, which corresponds to the $L1_0$-PtCo structure viewing from the *c*-axis (ESI Fig. S3†). The $L1_0$-PtCo structure can also be produced on the {100} surfaces perpendicular to the incident beam direction due to shape symmetry. Therefore, the particle is concluded to have $L1_0$-PtCo domains whose *c*-planes are parallel to the particle's {100} surfaces. The observed $L1_0$-PtCo domains suggest that the ratio of Pt:Co appears to be almost 1:1. Pt atoms have been reported to be segregated on the {100} surface of a Pt–Co alloy by annealing to minimize surface energy.[23] We think that Pt atoms segregate and form the {100} surface *via* diffusion, subsequently leading to the layer-by-layer formation of the $L1_0$-PtCo structure from the surface.

We also observed similar alternative stacks of Pt and Co layers in other nanoparticles when annealing at 700°C for 3 hours (See ESI Fig. S4†). The electron diffraction pattern taken at the corresponding region showed the diffraction rings of the fcc structure before annealing, while the pattern showed the diffraction rings of the ordered structure as well as ones of the fcc structure after the annealing (ESI Fig. S5†). The results are the evidence that the $L1_0$-PtCo multi-domain structure is reproducibly formed by annealing the Pt@Co core-shell particles at 700°C.

After 3 hours of annealing at 800°C, as shown in the HAADF image of Fig. 2e, we can see the periodic square pattern with darker contrast at the center and brighter contrast between the corner and the two edges, and this contrast is arranged throughout the particle. This indicates that the particle has a single domain of the $L1_2$-$Pt_3Co$ structure. However, we found that the contrast of the atomic columns was almost constant in 2–3 layers from the surface. They seemed to correspond to Pt atomic columns, although the contrast of these columns appears to be lower than one of the Pt atomic columns inside the nanoparticle due to the reduced thickness toward the edge. The experimental result indicates that the nanoparticle is a single $L1_2$ domain of Pt:Co~3:1 covered by Pt skin of ~0.5 nm in thickness. This configuration is consistent with the compositional ratio of Pt:Co = 4:1 of the particle overall.

To confirm the near-surface composition of the particle annealed at 700°C, EDS mapping was conducted (Fig. 3). The particle exhibits an $L1_0$ structure as well (ESI Fig. S4a†). The EDS results of the particle were presented in Figs. 3b and c, which indicated that a few Pt layers formed at the surface region of ~0.5 nm in depth. Contrary, Co atoms, that initially form the shell, are found around the near-surface but except for the top surface. The chemical distribution indicates that the diffusion proceeds enough to construct Pt shell and the near-surface consist of Pt and Co, but still insufficient to mix Pt and Co atoms in the particle overall. Fig. 3e illustrates an atomic resolution EDS map of the near-surface region. The two surface layers are made up of Pt atoms, while below the third layer, Pt(001) and Co(001) layers alternately occur. The EDS map is direct evidence of the $L1_0$ structure, in which Pt layers and Co layers are alternatively layered, for the annealed at 700°C nanoparticles.

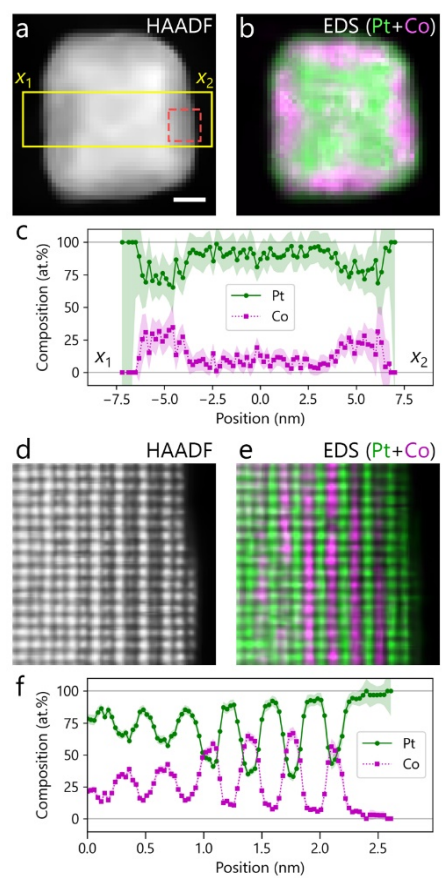

**Fig. 3** Analysis of a typical nanoparticle annealed at 700°C. (a) HAADF image. The scale bar corresponds to 3 nm. (b) EDS maps overlaid for Pt and Co intensities. (c) The composition profiles of Pt and Co as a function of the horizontal position on the line connecting $x_1$ and $x_2$. The EDS intensities of Pt and Co were averaged along the vertical direction in the rectangle region as shown in b and the averaged intensity profiles were converted into the composition profiles of Pt and Co. (d) Atom-resolved HAADF image of a sub-surface area, marked as the dashed square in a. (g) Atom-resolved EDS maps overlaid for Pt and Co intensities. (c) The composition profiles of Pt and Co as a function of the horizontal position. The EDS intensities in e were averaged along the vertical direction and converted into the composition profiles.

We found that the $L1_0$-PtCo structure was formed at the near-surface when annealing the Pt@Co core-shell nanoparticle at 700°C for 3 hours (Fig. 2c). This structure is unexpected because the $L1_2$-$Pt_3Co$ structure is more stable when the nanoparticles' composition ratio is Pt:Co = 4:1. At an annealing temperature of 800°C, the $L1_2$-$Pt_3Co$ structure was almost totally formed in the nanoparticles, and additional Pt atoms were segregated on the surface (Fig. 2d). Furthermore, even at 600°C, multiple growth nuclei of the $L1_2$-$Pt_3Co$ structure were seen to develop around the core-shell interface. As a result, the $L1_0$-PtCo structure at 700°C is thought to be produced *via* surface segregation of Pt atoms.

We think that such temperature-dependent ordered structures can be explained by a balance between surface segregation and diffusion. The diffusion length $\bar{L}$ (m) at temperature $T$ (K) can be estimated by the equation

$$\bar{L} = \sqrt{D_0 \tau} \exp\left(-\frac{Q}{2 k_B T}\right) \qquad (1).$$

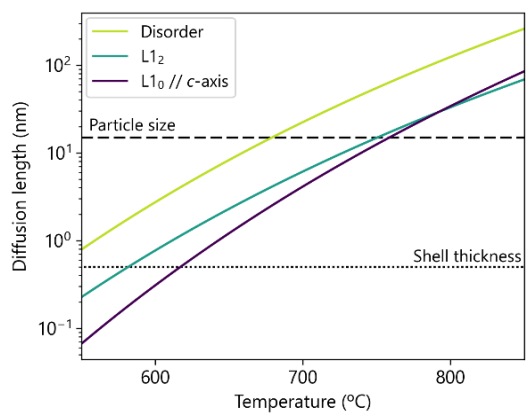

**Fig. 4** Diffusion lengths calculated from the parameters in previous reports. The plot shows interdiffusion in the disorder alloy, L1$_0$ structure along $c$-axes, and L1$_2$ structure. Note that Pt-Fe system was used.

Here, $D_0$ (m$^2$/sec) is pre-exponential factor, $Q$ (eV) is activation energy, $k_B$ = 8.62×10$^{-5}$ (eV/K) is Boltzmann constant, and $\tau$ = 10800 (sec) is annealing time, respectively. For evaluating the diffusion length, we used the diffusion parameters for Pt–Fe alloy, which seems to have a similar diffusion tendency (ESI Fig. S6†), because these values of Pt–Co alloy are not available to our knowledge. We assume that Pt and Fe atoms diffuse into each other (interdiffusion) to reduce the compositional difference. We calculated the diffusion coefficient and activation energy of interdiffusion in disordered alloys,[24] L1$_0$ along $c$-axis,[24] and L1$_2$.[25] ESI Table S1† summarizes the values of $D_0$ and $Q$.

Fig. 4 shows the diffusion lengths as a function of temperature. At 600°C, the diffusion length of L1$_2$ structure is comparable to the shell thickness. This is smaller than the diffusion length in the disordered alloy. The Pt or Co atom in the ordered structure must be changed to the position where another element should be. The diffusion in the ordered structures is generally unpreferable in terms of energy.[25] As a result, when the L1$_2$ structure is generated at the core-shell interface, the structure may prevent further diffusion. At 700°C, the diffusion length in the L1$_2$ structure is longer than the shell thickness. The diffusion length seems enough long to induce the surface segregation of Pt. However, the diffusion length is still smaller than the particle size so that the composition is not uniform in the particle overall. Considering the mutual diffusion between Co and Pt atoms in the particle, L1$_0$-PtCo structure seems more stable because of the compositional ratio of 1:1 near the surface locally. The diffusion length of Pt or Co atoms at 800°C is greater than the particle size. The Co atoms can completely move to the particle's core. The atoms can experience diffusion in the particle overall, resulting in the L1$_2$-Pt$_3$Co monodomain structure. The remained Pt atoms may be segregated on the surface.

In this study, when the annealing temperature and time are set so appropriately that the diffusion length is about the shell width, we find that the L1$_0$-PtCo structure can be formed around the near-surface of the Pt@Co core-shell nanoparticle. This near-surface structure is expected to be effective in promoting the oxygen reduction reaction. Because the L1$_0$-PtCo structure has a lower lattice constant,[8] the near-surface L1$_0$ structure is expected to introduce lattice contraction in the outermost Pt layer, which has been reported to optimize the energy level of $d$-orbital electrons for the oxygen reduction reaction.[26,27] We think that the ordered structure may be determined by the local compositional ratio at the surface area of Pt–Co alloy nanoparticles. If the compositional ratio of Pt and Co is almost equal, the L1$_0$-PtCo structure may be formed. However, if the Pt compositional ratio is high, the L1$_2$-Pt$_3$Co structure may occur. This consideration suggests that the distribution of components within the initial nanoparticles is important in regulating the near-surface structure of binary alloy nanoparticles.

## Conclusions

In summary, we find that the L1$_0$-PtCo structure can be formed at the near-surface of the Pt@Co core-shell nanoparticle when the annealing temperature and time are set so appropriately that the diffusion length is about the shell width. For the core-shell nanoparticle with a Co shell of around 0.5 nm width, the annealing temperature and time were 700°C and 3 hours, respectively. At 600°C, the diffusion length was predicted to be shorter than the shell width, resulting in the more stable L1$_2$-Pt$_3$Co structure being nucleated locally. At 800°C, the interdiffusion length was estimated to be longer than the shell width so that the L1$_2$-Pt$_3$Co structure was formed uniformly. The L1$_0$-PtCo structure at the near-surface is expected to be effective in promoting the oxygen reduction reaction. The advantage of core-shell nanoparticles is that the ordered structure can be controlled near the surface by regulating the annealing condition.

## Conflicts of interest

There are no conflicts to declare.

## Acknowledgements

The authors are indebted to Dr. Masaru Kato at STEM Co., Ltd., Japan for his contribution to the experimental setups. The authors are grateful to Mr. Koichi Higashimine at the Center for Nano Materials and Technology in Japan Advanced Institute of Science and Technology for technical supports. The authors also appreciate Dr. Masaki Kudo and Mr. Takaaki Toriyama at the Ultramicroscopy Research Center in Kyushu University for technical supports.

## References

1 Q. Wang, H. Tang, M. Wang, L. Guo, S. Chen and Z. Wei, *Chem. Commun.*, 2021, **57**, 4047–4050.
2 J. Bak, Y. Heo, T. G. Yun and S. Y. Chung, *ACS Nano*, 2020, **14**, 14323–14354.

3  D. Wang, H. L. Xin, R. Hovden, H. Wang, Y. Yu, D. A. Muller, F. J. Disalvo and H. D. Abruña, *Nat. Mater.*, 2013, **12**, 81–87.
4  J. Li, S. Sharma, X. Liu, Y.-T. Pan, J. S. Spendelow, M. Chi, Y. Jia, P. Zhang, D. A. Cullen, Z. Xi, H. Lin, Z. Yin, B. Shen, M. Muzzio, C. Yu, Y. S. Kim, A. A. Peterson, K. L. More, H. Zhu and S. Sun, *Joule*, 2019, **3**, 124–135.
5  Sun, Murray, Weller, Folks and Moser, *Science*, 2000, **287**, 1989–92.
6  D. J. Sellmyer, M. Yu and R. D. Kirby, *Nanostructured Mater.*, 1999, **12**, 1021–1026.
7  K. Yun, H. S. Nam and S. Kim, *Phys. Chem. Chem. Phys.*, 2020, **22**, 7787–7793.
8  P. Andreazza, V. Pierron-Bohnes, F. Tournus, C. Andreazza-Vignolle and V. Dupuis, *Surf. Sci. Rep.*, 2015, **70**, 188–258.
9  K. Hosoiri, F. Wang, S. Doi and T. Watanabe, *Mater. Trans.*, 2003, **44**, 653–656.
10  D. Bochicchio and R. Ferrando, *Phys. Rev. B - Condens. Matter Mater. Phys.*, 2013, **87**, 1–13.
11  U. Bardi, A. Atrei, P. N. Ross, E. Zanazzi and G. Rovida, *Surf. Sci.*, 1989, **211**–**212**, 441–447.
12  U. Bardi, A. Atrei, E. Zanazzi, G. Rovida and P. N. Ross, *Vacuum*, 1990, **41**, 437–440.
13  Y. Gauthier, P. Dolle, R. Baudoing-Savois, W. Hebenstreit, E. Platzgummer, M. Schmid and P. Varga, *Surf. Sci.*, 1998, **396**, 137–155.
14  Y. Gauthier, R. Baudoing-Savois, J. M. Bugnard, U. Bardi and A. Atrei, *Surf. Sci.*, 1992, **276**, 1–11.
15  Y. Gauthier, R. Baudoing-Savois, J. M. Bugnard, W. Hebenstreit, M. Schmid and P. Varga, *Surf. Sci.*, 2000, **466**, 155–166.
16  F. Li, Y. Zong, Y. Ma, M. Wang, W. Shang, P. Tao, C. Song, T. Deng, H. Zhu and J. Wu, *ACS Nano*, 2021, **15**, 5284–5293.
17  S.-I. Choi, R. Choi, S. W. Han and J. T. Park, *Chem. - A Eur. J.*, 2011, **17**, 12280–12284.
18  X. Xia, S. Xie, M. Liu, H. C. Peng, N. Lu, J. Wang, M. J. Kim and Y. Xia, *Proc. Natl. Acad. Sci. U. S. A.*, 2013, **110**, 6669–6673.
19  K. Aso, K. Shigematsu, T. Yamamoto and S. Matsumura, *Microscopy*, 2016, **65**, 391–399.
20  K. Aso, J. Maebe, X. Q. Tran, T. Yamamoto, Y. Oshima and S. Matsumura, *ACS Nano*, 2021, **15**, 12077–12085.
21  Y. Makinen, L. Azzari and A. Foi, in *2019 IEEE International Conference on Image Processing (ICIP)*, IEEE, 2019, pp. 185–189.
22  L. Azzari and A. Foi, *IEEE Signal Process. Lett.*, 2016, **23**, 1086–1090.
23  A. Christensen, A. Ruban, P. Stoltze, K. Jacobsen, H. Skriver, J. Nørskov and F. Besenbacher, *Phys. Rev. B - Condens. Matter Mater. Phys.*, 1997, **56**, 5822–5834.
24  A. Kushida, K. Tanaka and H. Numakura, *Mater. Trans.*, 2003, **44**, 59–62.
25  Y. Nos, T. Ikeda, H. Nakajima and H. Numakura, *Mater. Trans.*, 2003, **44**, 34–39.
26  J. Greeley, J. K. Nørskov and M. Mavrikakis, *Annu. Rev. Phys. Chem.*, 2002, **53**, 319–348.
27  B. T. Sneed, A. P. Young and C. K. Tsung, *Nanoscale*, 2015, **7**, 12248–12265.